\documentclass[preprint,showpacs,showkeys,groupedaddress,superscriptaddress]{revtex4}
\usepackage{epsf,epsfig}
\usepackage[psamsfonts]{amssymb}
\usepackage{amsmath}
\usepackage{bm}
\usepackage{natbib}
\begin{document}
\title{Quantum theory of a strongly-dissipative scalar field}
\author{Marjan Jafari}
\address{Department of Physics, Faculty of Science, Imam Khomeini International University, 34148 - 96818, Ghazvin, Iran}
\author{Fardin Kheirandish}
\address{Department of Physics, University of Kurdistan, P.O.Box 66177-15175, Sanandaj, Iran}

\begin{abstract}
The properties of a quantum dissipative scalar field is analyzed by Caldeira-Leggett model in strong-coupling regime. The Lagrangian of the total system is canonically quantized and the full Hamiltonian is diagonalized using Fano technique. A mode-dependent probability density is introduced. The steady state energy and correlation functions at finite temperature are calculated in terms of the probability density.
\end{abstract}


\vspace{2pc}


\maketitle
\section{Introduction}\label{Introduction}
\noindent Dissipative scalar field theories have an important place in quantum field theory and appear in various problems in physics. Even in quantum mechanics, introducing dissipation in the Hamiltonian of a quantum mechanical system, without invoking to a reservoir or a heat bath, is not a straightforward task and usually leads to inconsistencies in the formulation such as violation of Heisenberg uncertainty principle \cite{1-1,2-1,3-1,4-1,kheir}. There are two important approaches to take into account dissipation in quantum theory, namely phenomenological and canonical approach. In these approaches, the interaction between the system and its environment is defined such that an irreversible energy flow from system to its environment results \cite{1-061,2-061,4-3}. In the canonical approach, which we follow here, the whole system is described by a total Lagrangian. That is, the environment or reservoir is modeled by a collection of non-interacting harmonic oscillators with a continuum of frequencies. In this model, known as Caldeira-Leggett model, each oscillator in the reservoir is coupled to the main system through appropriate coupling functions \cite{7-11,32-B,33-B,34-B}. This model has been applied to a variety of dissipative quantum systems \cite{27-B}. In weak-coupling regime, one can apply Markovian approximation \cite{20-B} but in strong-coupling regime, non-Markovianity of the system should be taken into account \cite{44-B}. The quantum theory of a damped harmonic oscillator in strong coupling regime has been investigated in \cite{10-B,B}. 

In the present paper, following Cladeira-Leggett model, the environment or background of a scalar field is modeled by a continuum of harmonic oscillators and the quantum theory of the scalar field in the presence of such a medium is investigated in strong-coupling regime. We start from a total Lagrangian and quantize the total system in the framework of canonical quantization. The Hamiltonian of the total system is diagonalized using Fano diagonalization technique \cite{20-kh}. The main ingredient in this investigation is the appearance of a mode dependent probability density function describing the statistical properties of the system in its ground state. The quantum dynamics, energy and the correlation functions of dynamical variables related to the scalar field, are obtained in terms of the probability density function.

The layout of the paper is as follows: In Sec. 2, a classical Lagrangian is proposed for the total system and using Euler-Lagrange equations, the equations of motion for the scalar field and the reservoir are obtained. In Sec. 3, the system is canonically quantized and Hamiltonian is diagonalized using Fano diagonalization technique. In Sec. 4, the energy of the dissipative scalar field in ground state is obtained. In Sec. 5, the quantum dynamics of a dissipative scalar field is investigated and the correlation functions of dynamical variables are obtained in a thermal state.
\section{Classical dynamics }\label{Classical dynamics}
\noindent
Consider a real massive scalar field that is interacted with an environment defined by real scalar fields $X_\omega(x,t)$. We assume that the scalar field is defined in $1+1$-dimensional space-time and use natural units. The Lagrangian density for the field-environment system is
\begin{equation}
\mathcal{L}(t)=\mathcal{L}_s+\mathcal{L}_R+\mathcal{L}_{int}.
\end{equation}
The first term $\mathcal{L}_s$ is the Lagrangian of a massive scalar field or main system
\begin{equation}
\mathcal{L}_s  = \frac{1}{2} {\partial_\mu\phi(x,t)\partial^\mu\phi(x,t)}  - \frac{1}{2}m^2 \,\phi^2(x,t),
\end{equation}
the second term, is the Lagrangian density of a reservoir or heat bath consisting of a continuum of harmonic oscillators with displacements $\mathbf{X}_{\omega}$ and frequencies $\omega \in [0,\infty )$
\begin{equation}
\mathcal{L }_R= \frac{1}{2}\int\limits_0^\infty d\omega\,[\dot{X}_\omega ^2 (x,t) - \omega ^2 \,X_\omega ^2 (x,t)],
\end{equation}
and the last term is the interaction between the scalar field and its reservoir that is assumed to be linear and given by
\begin{equation}
\mathcal{L}_{\mbox{int}}  = {\int\limits_0^\infty  {d\omega } } f(\omega )\,\phi (x,t)\,X_{\omega} (x,t),
\end{equation}
where $f(\omega)$ is the coupling function between the scalar field and reservoir.
For simplicity we work in the reciprocal space and write the fields in terms of their spatial Fourier transforms
\begin{eqnarray}
  && \phi(x,t) = \int\limits_0^\infty \frac{dk}{2\pi}\,\left[e^{i k x}\,\phi (k,t)+e^{-i k x}\,\phi^* (k,t)\right], \\
  && X_\omega (x,t) = \int\limits_0^\infty \frac{dk}{2\pi}\,\left[e^{i k x}\,X_\omega (k,t)+e^{-i k x}\,X^*_\omega (k,t)\right].
\end{eqnarray}
where we have used the reality of the fields and restricted the range of the variable k to the half space $[0,\infty)$. In the reciprocal half space the Lagrangian components can be expressed as
\begin{equation}
\underline{L}_s=\int\limits_0^\infty \frac{dk}{\pi}\, \left(|\dot{\phi}(k,t)|^2-\omega_k^2 \,|\phi(k,t)|^2\right),
\end{equation}
where $\omega_k^2=(m^2+k^2)$.
\begin{equation}
\underline{L}_R=\int\limits_0^\infty \frac{dk}{\pi}\, \left(|\dot{X}_\omega(k,t)|^2-\omega^2 \,|X_\omega(k,t)|^2\right),
\end{equation}
\begin{equation}
\underline{L}_{\mbox{int}}=\int\limits_0^\infty \frac{dk}{\pi}\, \int \limits_0^\infty d\omega\, f(\omega)\left(\phi(k,t)\,X_\omega^*(k,t)+\phi^*(k,t)\,X_\omega(k,t)\right),
\end{equation}
where  $\phi^*(k,t)=\phi(-k,t)$ and $X_\omega^*(k,t)=X_\omega(-k,t)$. From Euler-Lagrange equations, the classical equations of motion for $\phi(k, t)$ and $X_\omega(k, t)$ are
\begin{eqnarray}\label{5}
&& (\partial_t^2+\omega_k^2)\,\phi(k,t)={\int\limits_0^\infty {d\omega}}\,f(\omega)\,X_\omega(k,t), \\
&& (\partial_t^2+\omega^2)\,X_\omega(k,t)=f(\omega)\,\phi(k,t).
\end{eqnarray}
\section{canonical quantization}\label{canonical quantization}
From the Lagrangian density $\mathcal{L}$ with measure $dk/\pi$
\begin{eqnarray}\label{Lagden}
  \mathcal{L} &=& \dot{\phi} (k,t)\dot{\phi}^* (k,t)+\int\limits_0^\infty d\omega\,\dot{X}_\omega (k,t) \dot{X}^*_\omega (k,t),\nonumber\\
  &-& \omega_k^2 \dot{\phi}(k,t)\dot{\phi}^*(k,t)-\int_0^\infty d\omega\, \omega^2 \dot{X}_\omega(k,t)\, \dot{X}^*_\omega(k,t)\nonumber\\
  &+& \int\limits_0^\infty d\omega\,[\phi (k,t) X^*_\omega (k,t)+\phi^* (k,t) X_\omega (k,t)],
\end{eqnarray}
the conjugate momenta corresponding to the fields $\phi$ and $X_\omega$ within the half k-space are defined by
\begin{equation}\label{6}
\pi(k,t)=\frac{\partial \mathcal{L}}{\partial \dot{\phi}^*}=\dot{\phi}(k,t),
\end{equation}
\begin{equation}\label{7}
\Pi_\omega(k,t)=\frac{\partial \mathcal{L}}{\partial{\dot{X}}^*_\omega}=\dot{X}_\omega(k,t).
\end{equation}
To quantize the theory canonically, the following equal-time commutation relations are be imposed on the fields and their conjugate momenta
\begin{equation}\label{8}
[\hat{\phi}^{\dag}(k,t),\hat{\pi}(k',t)]=i\,\delta(k-k'),
\end{equation}
\begin{equation}\label{9}
[\hat{X}_\omega^{\dag}(k,t),\hat{\Pi}_{\omega'}(k',t)]=i\,\delta(k-k')\delta(\omega-\omega'),
\end{equation}
with all other equal-time commutators being zero. Using the canonical momenta and the Lagrangian density, we find the Hamiltonian density of the field-environment system as follows
\begin{eqnarray}\label{10}
\underline{\mathcal{H}} &=&
\left[|\hat{\pi}(k,t)|^2+\omega_k^2\,|\hat{\phi}(k,t)|^2\right]+{\int\limits_0^\infty} d\omega\,\left(|\hat{\Pi}_\omega(k,t)|^2 +\omega^2\,|\hat{X}_\omega(k,t)|^2\right)\nonumber\\
&-& {\int\limits_0^\infty}{d\omega}\,f(\omega) [\hat{\phi}^{\dag}(k,t)\,\hat{X}_\omega(k,t)+\hat{\phi}(k,t)\,\hat{X}_\omega^{\dag}(k,t)].
\end{eqnarray}
We can rewrite the Hamiltonian density in a minimal-coupling form
\begin{eqnarray}\label{11}
\underline{\mathcal{H}} &=& \left[|\hat{\pi}(k,t)|^2+\Omega_k^2\,|\hat{\phi}(k,t)|^2\right]\nonumber\\
&+& {\int\limits_0^\infty}{d\omega}\,\left[|\hat{\Pi}_\omega(k,t)|^2 +\omega^2\,|\hat{X}_\omega(k,t)-\frac{f(\omega)}{\omega^2}\,\hat{\phi}(k,t)|^2\right],
\end{eqnarray}
where
\begin{equation}\label{12}
\Omega^2_k=\omega_k^2-\int\limits_0^\infty d\omega \,\frac{f^2(\omega)}{\omega^2}.
\end{equation}
The Hamiltonian is positive only if this quantity is positive. The positivity of the Hamiltonian induce a physical restriction on the strength of the coupling
\begin{equation}\label{13}
\omega_k^2 > \int\limits_0^\infty d\omega\,\frac{f^2(\omega)}{\omega^2}.
\end{equation}
To facilitate the calculations, let us introduce the following annihilation operators
\begin{eqnarray}\label{121}
\hat{a}_k (t) &=& \sqrt{\frac{\omega_k}{2}}\,\left[ \hat{\phi}(k,t)+\frac{i}{\omega_k}\,\hat{\pi} (k,t)\right], \\ \label{121a}
\hat{a}^{\dag}_k (t) &=& \sqrt{\frac{\omega_k}{2}}\,\left[ \hat{\phi}^{\dag} (k,t)-\frac{i}{\omega_k}\,\hat{\pi}^{\dag} (k,t)\right], \\\label{121b}
\hat{b}_k (\omega,t) &=& \sqrt{\frac{\omega}{2}}\,\left[ \hat{X}_\omega (k,t)+\frac{i}{\omega}\,\hat{\Pi}_\omega (k,t)\right],\\\label{121c}
\hat{b}^{\dag}_k (\omega,t) &=& \sqrt{\frac{\omega}{2}}\,\left[ \hat{X}^{\dag}_\omega (k,t)-\frac{i}{\omega}\,\hat{\Pi}^{\dag}_\omega (k,t)\right],
\end{eqnarray}
satisfying bosonic equal-time commutation relations
\begin{eqnarray}\label{ladderCommut}\label{122}
  && [\hat{a}_k (t), \hat{a}_{k'} (t)]=[\hat{a}^{\dag}_k (t), \hat{a}^{\dag}_{k'} (t)]=0,\nonumber\\
  && [\hat{a}_k (t), \hat{a}^{\dag}_{k'} (t)]=\delta(k-k'),\\\label{122a}
  && [\hat{b}_k (\omega,t), \hat{b}_{k'} (\omega',t)]=[\hat{b}^{\dag}_k (\omega,t), \hat{b}^{\dag}_{k'} (\omega',t)]=0,\nonumber\\
  && [\hat{b}_k (\omega,t), \hat{b}^{\dag}_{k'} (\omega',t)]=\delta(k-k')\,\delta(\omega-\omega').
\end{eqnarray}
The commutation relations (\ref{122},\ref{122a}) in contrast to the previous relations (\ref{8},\ref{9}), which
were correct only in the half k-space, are now valid in the whole reciprocal space.
Using the equations (\ref{121}-\ref{121c}), we can write the canonical variables $\phi$ and $X_\omega$ in terms of the creation and annihilation operators as
\begin{eqnarray}\label{expand}
\hat{\phi} (k,t) &=& \frac{1}{\sqrt{2 \omega_k}}\left[\hat{a}_k (t)+\hat{a}^{\dag}_{-k} (t) \right], \\
\hat{\pi} (k,t) &=& -i\sqrt{\frac{\omega_k}{2}}\left[\hat{a}_{k} (t) -\hat{a}^{\dag}_{-k} (t)\right], \\
\hat{X}_\omega (k,t) &=& \frac{1}{\sqrt{2 \omega}}\left[\hat{b}_k (\omega,t)+\hat{b}^{\dag}_{-k} (\omega,t)\right], \\
\hat{ \Pi}_\omega (k,t) &=& -i\sqrt{\frac{\omega}{2}}\left[\hat{b}_{k} (\omega,t)-\hat{b}^{\dag}_{-k} (\omega,t)\right].
\end{eqnarray}
Hamiltonian in terms of creation and annihilation operators is
\begin{eqnarray}\label{Hamil}
 H &=& \int \limits_{-\infty}^\infty \frac{dk}{\pi}\,\omega_k \,\hat{a}^{\dag}_k (t) \hat{a}_k (t) + \int \limits_{-\infty}^\infty \frac{dk}{\pi}\,\int \limits_0^\infty d\omega\,\omega\,\hat{b}^{\dag}_k (\omega,t) \hat{b}_k (\omega,t)\nonumber\\
 &-& \int \limits_{-\infty}^\infty \frac{dk}{\pi}\,\int_0^\infty d\omega\,v(\omega,k)\,[a^{\dag}_{k} (t)+\hat{a}_{-k} (t)][
 \hat{b}^{\dag}_{-k} (\omega,t) +\hat{b}_k (\omega,t)],
\end{eqnarray}
where the coupling is given by $v(\omega,k)=f(\omega)/2\sqrt{\omega \omega_k}$.
\subsection{Hamiltonian diagonalization}\label{Hamiltonian diagonalisation}
Hamiltonian can be diagonalized by introducing a complete set of eigenoperators $\hat{C}_k (\omega,t)$ and $\hat{C}^\dag_k (\omega,t)$, obeying bosonic commutation relation
\begin{equation}\label{22}
\left[\hat{C}_k(\omega,t),\hat{C}_{k'}^\dag(\omega',t)\right]=\delta(\omega-\omega')\delta(k-k').
\end{equation}
These operators satisfy the equations
\begin{eqnarray}\label{23}
&& \left[\hat{C}_k (\omega,t), \hat{H}\right]=\omega\,\hat{C}_k (\omega,t),\\
&& \left[\hat{C}^\dag_k (\omega,t), \hat{H}\right]=-\omega\,\hat{C}^\dag_k (\omega,t),\\
&& \hat{H}=\int\limits_{-\infty}^\infty \frac{dk}{\pi}\,\int\limits_0^\infty d\omega\,\omega\,\hat{C}^{\dag}_k (\omega,t)\hat{C}_k (\omega,t).
\end{eqnarray}
From now on, for notational convenience, we do not write the explicit time dependence of operators. The eigen-operators can be written as a superposition of a complete set of scalar field and reservoir operators
\begin{eqnarray}\label{24}
\hat{C}_k (\omega) &=& \alpha_k (\omega)\hat{a}_k+\beta_k(\omega)\hat{a}^\dag_{-k} \nonumber\\
&+& \int\limits_0^\infty {d\omega'}\left[\gamma_{k}(\omega,\omega')\hat{b}_k (\omega')+\delta_{k}(\omega,\omega')\hat{b}^\dag_{-k}(\omega')\right],
\end{eqnarray}
by substituting (\ref{24}) in (\ref{23}) we find the coefficients as
\begin{equation}\label{28}
\beta_k(\omega)=\frac{\omega-\omega_k}{\omega+\omega_k}\,\alpha_k (\omega),
\end{equation}
\begin{equation}\label{29}
\delta_k(\omega,\omega')=\left(\frac{1}{\omega+\omega'}\right)\,v(\omega',k)\,\frac{-2\omega_k}{\omega+\omega_k}\alpha_k(\omega)
\end{equation}
and
\begin{equation}\label{30}
\gamma_k (\omega,\omega')=\left(\frac{P}{\omega-\omega'}+Y_k(\omega)\delta(\omega-\omega') \right)v(\omega',k)\frac{-2\omega_k}{\omega+\omega_k}\alpha_k(\omega),
\end{equation}
where
\begin{equation}\label{31}
Y_k(\omega)=\frac{1}{v^2(\omega,k)}\left[\frac{(\omega_k^2-\omega^2)}{2\omega_k}-\int\limits_0^\infty{d\omega'}\left(\frac{P}{\omega-\omega'}-\frac{1}{\omega+\omega'} \right)v^2(\omega',k) \right]
\end{equation}
If we insert $\hat{C}_k(\omega)$, expressed in terms of $\alpha_k(\omega)$, into the commutation relation (\ref{22}), we find
\begin{equation}
| {\alpha_k(\omega)} |^2=\frac{(\omega+\omega_k)^2}{4\omega_k^2 v^2(\omega,k)}\left(\frac{1}{Y_k^2(\omega)+\pi^2} \right).
\end{equation}
The field operators in terms of eigen-operators are as follows
\begin{eqnarray}
  \hat{a}_k=\int \limits_0^\infty d\omega \left(\alpha_k^*(\omega)\hat{C}_k(\omega)-\beta_k(\omega)\hat{C}_{-k}^\dag(\omega) \right)\\
  \hat{a}_k^\dag=\int \limits_0^\infty d\omega \left(\alpha_k(\omega)\hat{C}_k^\dag (\omega)-\beta_k^*(\omega)\hat{C}_{-k}(\omega) \right)
 \end{eqnarray}
The requirement that these operators satisfy the familiar boson commutation relation, $[\hat{a}_k,\hat{a}_{k'}^\dag]=\delta(k-k')$, leads to
\begin{equation}\label{40}
\int\limits_0^\infty d\omega \left[| {\alpha_k(\omega)} |^2-| {\beta_k(\omega)} |^2 \right]=\int\limits_0^\infty d\omega | {\alpha_k(\omega)} |^2\frac{4\omega_k\omega}{(\omega+\omega_k)^2}=1.
\end{equation}
The integrand in (\ref{40}) is positive for all frequencies and hence has the mathematical form of a frequency probability distribution
\begin{equation}
P_k(\omega)=| {\alpha_k(\omega)} |^2\frac{4\omega_k\omega}{(\omega+\omega_k)^2}.
\end{equation}
So having this distribution, we can find the average value of an arbitrary function of frequency
\begin{equation}
{\langle {\langle {f(\omega)} \rangle } \rangle}_k=\int\limits_0^\infty d\omega \,f(\omega)\,P_k(\omega).
\end{equation}
\section{Ground-state}
The ground-state $|0\rangle$ of the eigenoperators $\hat{C}_k(\omega)$ is defined by
\begin{equation}
\hat{C}_k(\omega)|0\rangle=0,
\end{equation}
the ground state of the scalar field can be obtained by tracing out the environmental degree of freedom
\begin{equation}
\hat{\rho}_s=\mbox{tr}_R(|0\rangle \langle 0|).
\end{equation}
Following \cite{BB}, we use the characteristic function to determine the form of the mixed state
\begin{equation}\label{kai}
\chi(k,\eta)=\mbox{tr}[\rho\, \exp(\eta \hat{a}_{-k}^{\dag}-\eta^*\,\hat{a}_k)]=\langle 0 | \exp(\eta \hat{a}_{-k}^\dag-\eta^*\hat{a}_k)|0\rangle.
\end{equation}
By inserting $\hat{a}_k$ and $\hat{a}_k^\dag$ in terms of the eigenoperators into (\ref{kai}) we find
\begin{equation}
\chi(k,\eta)=\exp{\left[-\frac{1}{2}\int\limits_0^\infty d\omega {|\eta \alpha_k(\omega)+\eta^* \beta_k(\omega) |}^2 \right]},
\end{equation}
which can be rewritten the charactristic function in terms of probability density
\begin{eqnarray}
\chi(k,\eta)=\exp\left[-\frac{1}{2}\int\limits_0^\infty P_k(\omega) \left(\frac{\omega}{\omega_k}\eta_r^2+\frac{\omega_k}{\omega}\eta_i^2 \right) \right],\\
=\exp\left[-\frac{1}{2}\left(\frac{{\langle {\langle {\omega} \rangle } \rangle}_k}{\omega_k}\eta_r^2+{\langle {\langle {\omega^{-1}} \rangle } \rangle}_k \omega_k \eta_i^2 \right) \right],
\end{eqnarray}
where $\eta_i$ and $\eta_r$ are imaginary and real part of $\eta$. The following relations can be easily obtained
\begin{eqnarray}
\langle 0 |\hat{\phi}(k,0)| 0 \rangle=0,\\
\langle 0 |\hat{\pi}(k,0)| 0 \rangle=0,\\
\langle 0 |\hat{\phi}^2(k,0)| 0 \rangle={{\langle {\langle {\omega^{-1}} \rangle } \rangle}_k}/{2},\\
\langle 0 |\hat{\pi}^2(k,0)| 0 \rangle={{\langle {\langle {\omega} \rangle } \rangle}_k}/{2}.
\end{eqnarray}
The mean energy of the scalar field for $k$-mode is
\begin{equation}
\frac{1}{2}\langle \pi^2(k,0)+(k^2+m^2)\phi^2(k,0)\rangle=\frac{\omega_k}{4}\left(\frac{{\langle {\langle {\omega} \rangle } \rangle}_k}{\omega_k}+\omega_k {\langle {\langle {\omega^{-1}}\rangle } \rangle}_k \right)\, >\frac{\omega_k}{2},
\end{equation}
which is more than the ground state energy of the free mode due to the amount of energy to be paid to decouple the scalar field from bath \cite{9-B}.
The mean-square value of the scalar field in space-time is given by
\begin{eqnarray}
\langle 0 |\hat{\phi}^2(x,t)| 0 \rangle=\int dk \frac{1}{2\pi^2\omega_k}\left[\frac{{\langle {\langle {\omega} \rangle } \rangle}_k}{2\omega_k} -\cos^2(kx)\left( \frac{1}{\omega_k}-{\langle {\langle {\omega^{-1}}\rangle }\rangle}_k\right)\right].
\end{eqnarray}
\section{Scalar field dynamics}
From the diagonalised Hamiltonian it is straightforward to evaluate the time-evolution of the scalar field. The time-evolution of eigenoperators are
\begin{eqnarray}
\hat{C}_k(\omega,t) &=& \hat{C}_k(\omega,0)e^{-i\omega t},\\
\hat{C}_k^\dag(\omega,t) &=& \hat{C}_k^\dag(\omega,0)e^{i\omega t}.
\end{eqnarray}
Using these expressions, we find the time-evolution of the $k$-mode annihilation operator of the scalar field
\begin{eqnarray}
\hat{a}_k(t) &=& \int\limits_0^\infty d\omega\,[\alpha_k^*(\omega)\hat{C}_k(\omega,0)e^{-i\omega t}-\beta_k(\omega)\hat{C}_{-k}^\dag(\omega,0)e^{i\omega t}],\nonumber\\
&=& \int\limits _0 ^\infty d\omega  \alpha_k^*(\omega) e^{-i\omega t}\{\alpha_k(\omega)\,\hat{a}_k(0)+\beta_k(\omega)\,\hat{a}_{-k}^\dag(0)\nonumber\\
&+& \int\limits_0^\infty d{\omega'}(\gamma_k(\omega,\omega')\,\hat{b}_k(\omega',0)+\delta_k(\omega,\omega')\,\hat{b}_{-k}^\dag(\omega',0)) \}\nonumber\\
&-& \int\limits _0 ^\infty d\omega \beta_k(\omega)e^{i\omega t}\{\alpha_k^*(\omega)\,\hat{a}_k^\dag(0)+\beta_k^*(\omega)\,\hat{a}_{-k}(0)\nonumber\\
&+& \int \limits _0 ^\infty d\omega'\,(\gamma_k^*(\omega,\omega')\,\hat{b}_k^\dag(\omega' ,0)+\delta_k(\omega,\omega')\,\hat{b}_{-k}(\omega' ,0)) \}.
\end{eqnarray}
When the environment is a stationary state, we have $\langle \hat{b}_k(\omega,0)\rangle=\langle \hat{b}_k^\dag(\omega,0)\rangle=0$, and the expectation values of the position and momentum operators are
\begin{eqnarray}
\langle\hat{\phi}(x,t)\rangle &=& \langle\langle\cos(\omega t)\rangle\rangle_k\langle\hat{\phi}(x,0)\rangle+\frac{1}{2\omega_k}\langle\langle\omega^{-1}\sin(\omega t)\rangle\rangle_k\langle\hat{\pi}(x,0)\rangle,\\
\langle\hat{\pi}(x,t)\rangle &=& \langle\langle\cos(\omega t)\rangle\rangle_k\langle\hat{\pi}(x,0)\rangle -\frac{\omega_k}{2}\langle\langle\omega^{-1}\sin(\omega t)\rangle\rangle_k\langle\hat{\phi}(x,0)\rangle.
\end{eqnarray}
In thermal equilibrium, we have
\begin{equation}\label{CC}
\langle \hat{C}_k^\dag(\omega,0)\hat{C}_{k'}(\omega',0)\rangle=N(\omega)\delta(k-k')\delta(\omega-\omega'),
\end{equation}
where
\begin{equation}\label{N}
N(\omega)=\frac{1}{\exp(\hbar\omega/k_B T)-1}.
\end{equation}
The expectation value of $k$-mode number operator of the scalar field in a thermal state is
\begin{eqnarray}
\langle\hat{a}^\dag_k(t)\hat{a}_k(t)\rangle_{\mbox{th}} &=&\frac{1}{2\omega_k}\langle\langle\omega N(\omega)\rangle\rangle_k+\frac{\omega_k}{2}\langle\langle\frac{N(\omega)}{\omega}\rangle\rangle_k+
\frac{\langle\langle\omega\rangle\rangle_k}{4\omega_k}+\frac{\omega_k}{4}\frac{1}{\langle\langle\omega\rangle\rangle_k}-\frac{1}{2},
\end{eqnarray}
which consists of zero and finite temperature contributions.
\section{Conclusion}
The quantum theory of a dissipative scalar field in strong-coupling regime and in the framework of canonical quantization investigated. The Hamiltonian of the system determined and diagonalized using Fano technique. A probability function in frequency describing the statistical properties of the scalar field in the presence of a dissipative medium was found. Explicit expressions for time-evolution of dynamical variables and their correlation functions at finite temperature were determined.
\section*{References}

\end{document}